# Fermi Surface Nesting and Nanoscale Fluctuating Charge/Orbital Ordering in Colossal Magnetoresistive Oxides

Y. -D. Chuang[1,2], A. D. Gromko[1], D. S. Dessau[1], T. Kimura[3], Y. Tokura[3,4]

1. Department of Physics, University of Colorado, Boulder, Colorado, 80309-0390

2. Advanced Light Source (ALS), Lawrence Berkeley National Laboratory, Berkeley, CA 94720

3. Department of Applied Physics, University of Tokyo, Tokyo 113-8656, Japan

4. Joint Research Center for Atom Technology (JRCAT), Tsukuba 305-0046, Japan

**ABSTRACT**

We used high resolution angle-resolved photoemission spectroscopy to reveal the Fermi surface and key transport parameters of the metallic state of the layered Colossal Magnetoresistive (CMR) oxide $La_{1.2}Sr_{1.8}Mn_2O_7$. With these parameters the calculated in-plane conductivity is nearly one order of magnitude larger than the measured DC conductivity. This discrepancy can be accounted for by including the pseudogap which removes at least 90% of the spectral weight at the Fermi energy. Key to the pseudogap and many other properties are the parallel straight Fermi surface sections which are highly susceptible to nesting instabilities. These nesting instabilities produce nanoscale fluctuating charge/orbital modulations which cooperate with Jahn-Teller distortions and compete with the electron itinerancy favored by double exchange.

**INTRODUCTION**

Many central topics in condensed matter physics are related to the competition between various types of order, with this competition often resulting in nanoscale phase separation and complex phase diagrams. For example, high temperature superconductivity in the doped copper oxides (cuprates) occurs in proximity to an antiferromagnetically ordered ground state, as well as to a mixed state with one-dimensional (1D) nanoscale charge "stripe" order. There has been much debate about how these states may coexist, compete or even cooperate with each other *(1)*. However, uncovering the details has been a difficult task at best, mostly because in the important regimes the various ordering phenomena fluctuate wildly both in time and space. We show that similar competition/cooperation behavior and nanoscale phase separation are also critical in the colossal magnetoresistive (CMR) oxides, and should be considered as a hallmark of modern condensed matter physics.

We have come to this conclusion on the basis of angle resolved photoemission (ARPES) experiments, which directly tell us the $\vec{k}$ dependant electronic structure of these materials. These experiments enable us to make a direct determination of the key transport parameters of the manganites, including the Fermi surface (FS) topology, mean free path $\lambda$, Fermi velocity $v_F$, effective mass m* and number of carriers *n*. In addition, our work highlights the importance of the pseudogap. By connecting with recent x-ray scattering experiments, we propose a picture explaining the origins of the pseudogap and the connection between various mechanisms in the CMR effect.

**MATERIALS SYSTEM**

We studied single crystals of the bilayer family of manganites. Each Mn atom sits in the center of an octahedron with an O atom at each of the six corners. Two such octahedra share one apical O atom and form a biplane where the CMR effect is believed to occur. Between each biplane, there are La/Sr atoms weakly binding the crystal together. The cleavage in this La/Sr rock-salt layer produces mirror like surfaces without dangling bonds or any known surface reconstruction.

The material we studied has the chemical composition $La_{1.2}Sr_{1.8}Mn_2O_7$ corresponding to a doping level x=0.4 (0.4 holes per Mn site) in the doping phase diagram (Figure 1A) *(2)*. There are many types of magnetic ordering phenomena observed at low temperatures, implying a close proximity in energy scales. Charge and orbital orderings have also been observed at the doping levels x=0.4 and 0.5 *(3,4,5)* and will be discussed in detail later in the paper. The CMR effect is observed between the doping levels x ~ 0.3 to 0.5 near the transition from the high temperature paramagnetic insulating phase to the low temperature ferromagnetic metallic phase. For example, Fig 1(B) shows the resistivity versus temperature curve for the doping level x=0.4 *(6)*, corresponding to the samples we studied. Application of a magnetic field drives the system toward the metallic state, producing the negative CMR effect near the magnetic transition temperature $T_c$ ~ 120K. The connection between the metal-insulator and the magnetic transitions can be understood to first order within the double exchange theory *(7,8,9)*, although it is also appreciated that double-exchange alone can only account for a roughly 30% change in conductivity *(10)*, whereas the observed effect may be several orders of magnitude. Hence the community has been very actively studying these compounds in the hope of uncovering new physics to explain the CMR effect *(11)*.

**TECHNIQUE**

We used high resolution ARPES to uncover the details of the FS and $\vec{k}$ dependent electronic structure. In an ARPES experiment, a monochromatic photon beam ejects electrons which are then collected and energy and momentum are analyzed. The number of photoelectrons with particular final state kinetic energy $E_f$ is closely related to the number of electrons with the initial state energy $E_i = E_f - h\nu - \phi$, where $h\nu$ is the known energy of the exciting photon and $\phi$ is the work function. Under the standard approximation, ARPES directly provides the information about the momentum-resolved single-particle spectral function $A(\vec{k},\omega)$, which when integrated over all momenta gives the density of electronic states. ARPES has been widely used in the study of high temperature superconductors and has proven to be a very powerful tool for probing the electronic structure*(12)*. However, there haven't been many ARPES results on the CMR materials because of very weak near-Fermi emission and limited resolution. By using the new undulator beamline 10 and the High Energy Resolution Spectrometer (HERS) endstation at the Advanced Light Source (ALS) of the Lawrence Berkeley National Laboratory, we were able to overcome these limitations *(13)*.

**RESULTS- FS**

The near Fermi energy ($E_F$) spectral weight measured throughout the Brillouin zone (Fig 2A) gives an experimental mapping of the 2D FS. This intensity plot was obtained by first normalizing roughly 600 individual energy distribution curves (EDCs)( intensity versus binding energies relative to $E_F$ at fixed emission angle) with high harmonic emission above $E_F$ *(14)* and then integrating the spectral weight within ±200 meV of $E_F$. The raw data were taken only in the lower half of the Brillouin zone but were reflected about the $(0,0)-(\pi,0)$ line based on symmetry arguments. We overlay this intensity plot with the FS topology predicted by a spin-polarized local density approximation (LSDA) calculation *(15,16)*. This theoretical FS consists mainly of two parts: one small electron pocket centered around the zone center $(0,0)$ which has predominantly out-of-plane $d_{3z^2-r^2}$ orbital character, and two concentric large hole pockets centered around the zone corners $(\pm\pi,\pm\pi)$ which have predominantly in-plane $d_{x^2-y^2}$ orbital character. The experimental image plot matches the theoretical FS relatively well: The higher intensity around $(0,0)$ corresponds to the small electron pocket and the four L-shaped loci around the corners correspond to portions of the hole pockets. Because of the nearly out-of-plane photon polarization, the emission around $(0,0)$ is greatly enhanced. The small asymmetry in the intensity along the $(0,0)-(\pi,0)$ line is also likely to be a polarization effect, but slight instrumental asymmetries may contribute to this as well. An important experimental observation is that the predicted splitting of the two hole-like pieces, which theoretically is due to the coupling between two MnO$_2$ planes per unit cell, is not observed indicating the strong in-plane correlation effects which reduce the theoretical splitting below detectable limits. A similar situation has been observed in high temperature superconductors such as Bi$_2$Sr$_2$CaCu$_2$O$_{8+\delta}$, in which the splitting predicted by band theory *(17)* is reduced by at least a factor of 3, with the splitting escaping detection until this year *(18,19)*, and then only in the most heavily overdoped compounds.

**E versus k RELATIONS**

More details of the electronic structure can be obtained through the study of energy ($E$) versus momentum ($\vec{k}$) dispersion relations. The angular slice shown in Fig 2B, taken at the 10° φ angle and indicated by the white line in Fig 2A, shows the photoemission intensity (denoted by the false color scale in which yellow = high intensity) as a function of binding energy relative to $E_F$. There clearly exists a dispersive feature of roughly parabolic shape with maximum spectral weight around binding energy 1 eV and θ = 0°. As this peak disperses toward $E_F$, it loses spectral weight until it is no longer visible below roughly 0.3 eV. However, if we blow up the spectra below 0.2 eV and rescale the colors (Fig 2C) we see that the weight does continue all the way to $E_F$, although it is dropping dramatically. It is this very low weight remaining at $E_F$ that is responsible for the FS plot displayed in Fig 2A.

Typical EDCs obtained by taking constant-angle cuts from Fig 2B and Fig 2C are shown in Fig 2D and Fig 2E respectively. Every third EDC is presented, with the corresponding angles listed in Fig 2E. The EDCs shown in Fig 2D are essentially identical to those presented previously *(16,20)*; however the greatly improved energy and angular resolution in the current study allow us for the first time to clearly resolve the spectral weight at $E_F$. This can be seen around emission angles $\pm 4.7^O$ in Fig 2E. When these data are overlain with the Au spectrum taken at the same temperature (Fig 2F), the almost perfect match indicates that the step at $E_F$ is simply due to the Fermi-Dirac distribution which provides the direct evidence for the existence of a FS in this system.

**MDC'S AND KEY TRANSPORT PARAMETERS**

Instead of using the EDCs, which are broad and have unusual line shape and background, we can extract much more information by analyzing the data in terms of momentum distribution curves (MDCs: intensity versus emission angles at fixed binding energy). Figure 2G shows a series of MDCs from Fig 2C. We see that the MDCs have much simpler line shape which can easily be fitted with a double Lorentzian (or nearly as good, a Gaussian) function for the two branches of the dispersion on top of a roughly constant background. The fitting gives us the dispersion relation $E(\vec{k})$, the peak width $\Delta k$ (Fig 3B) and the spectral weight under the peak (Fig 3A) for each MDC, which can then be plotted against the binding energies.

The results in Fig 3B allow us to extract the key transport parameters for these in-plane states, which we find to be essentially independent of the location along the roughly 1D FS(21). The Fermi velocity $v_F$ determined by the slope of the dispersion at $E_F$, $v_F = \frac{1}{\hbar}\frac{dE}{d\vec{k}}$, is approximately 0.038 $c$ ($c$ is the speed of light). The effective mass $m^*$, from fitting the experimental dispersion relation $E(\vec{k})$ with a parabola, is approximately 0.27 $m_e$ ($m_e$ is the free electron mass), which is within ~10% of the band theory result for these states. This light mass is surprising for this system.

The peak width at $E_F$ is around 1.25° half-width half-maximum, or equivalently the momentum width is $0.09\frac{\pi}{a} \approx 0.07 \overset{o}{A}^{-1}$ with $a$ the in-plane lattice constant. Because our angular resolution is significantly better than this, we consider it to be intrinsic and originating from the finite mean free path $\lambda = \frac{1}{\Delta k} \approx 14 \overset{o}{A}$. This mean free path is roughly 7 times the in-plane spacing between the Mn and O atoms. A similar value has been estimated by Li *et al.* from the magnetic-field dependence of the in-plane conductivity, *(22)* which they fit to the theoretical form of quantum interference effects suggested by Abrikosov*(23)*. We also can calculate the mean free time between scattering events $\tau = \frac{\lambda}{v_F}$, obtaining a value of approximately 1.24 fs. Finally, the estimated number of in-plane carriers ($n$) from the volume of the FS pockets centered around $(\pi,\pi)$, which cover nearly half of the full Brillioun zone size, gives $n \approx 3.4*10^{21} holes/cm^3$.

From these parameters the in-plane dc conductivity can be estimated on the basis of the Drude formula: $\sigma = \frac{ne^2\tau}{m^*}$, where we note that all the parameters are approximately independent of the location on the in-plane FS. Using these numbers we obtain the resistivity $\rho_{ARPES} \approx 2*10^{-4} \Omega \bullet cm$ whereas the measured low-temperature resistivity $\rho_o \approx 2*10^{-3} \Omega \bullet cm$ (see Fig 1B) is nearly one order of magnitude larger than $\rho_{ARPES}$. The low temperature state of these materials is a very poor metal with the resistivity on the order of the inverse of Mott's minimum metallic conductivity. Such poor conductivity would typically be regarded as the outcome of either very strong scattering (a short mean free path on the order of the lattice constant) or a very small number of carriers (small FS volume), although our data directly show that neither can be the case.

**DISCUSSION-PSEUDOGAP**

In order to resolve this inconsistency, other physics must be included, the most obvious of which is the suppression of spectral weight near $E_F$ (Fig 3A). Unusual behavior in the spectral weight of the dispersive peak, falling rapidly over quite a large energy scale (nearly 0.5 eV) as the peak approaches $E_F$ (Fig 3A), was previously termed the ``pseudogap''.*(16,20)* It is not expected in the simple theory of metals, in which the spectral weight remains constant until it is cut by the Fermi function (black dotted line). This pseudogap will decrease the conductivity by removing a large portion of the carriers from the conduction process. As the temperature is raised above $T_c$ (Fig 3C) the pseudogap removes nearly all remaining spectral weight at $E_F$, which is consistent with the insulating behavior of the high temperature phase. Therefore, the pseudogap appears to be critical for explaining the poor conductivity of both phases, and presumably should be important for explaining the transition between them as well.

A number of possibilities exist to explain the pseudogap, both intrinsic and extrinsic. We begin with the extrinsic possibilities which we argue can be largely excluded. First, there is the extrinsic ohmic loss effect due to poor conductivity*(24)*. However recent experiments *(25,26)* and theories *(26)* convincingly argue that this is very unlikely to be a major concern. Next, there is the matrix element effect which will modulate the photoemission spectral intensities and in general has both energy and momentum-dependent terms. To guard against this, we have taken spectra at various photon energies (between ~ 20 and 50 eV) and polarizations. Although the weights do vary slightly with photon energy, the weight loss trend is still robust. Additionally, optical conductivity experiments have also shown the absence of the Drude peak in the same system (27,28) which we believe to be closely related to the pseudogap observed in ARPES.

Intrinsic possibilities to explain the pseudogap that have been discussed in the literature include Jahn Teller effects *(29,30,31)*, polaronic*(10,32)* or bipolaronic *(33)* effects, strong on-site Coulomb interactions *(34)*, and electronic phase separation*(35,36)*. Whatever the potential importance of these ideas, there is a new piece of evidence that we feel gives a very strong clue towards a new explanation for the pseudogap and a new picture to incorporate these mechanisms.

**FS NESTING AND charge density wave (CDW) INSTABILITIES**

This new evidence lies in the FS topology presented in Fig 2A. The hole-like portions are very straight (probably more so than expected from the LSDA calculation, especially near the corners of the hole pockets). Such straight parallel segments indicate that the physics of these compounds should be considered quasi 1D. This is consistent with recent experimental*(37)* and theoretical works*(38)* showing stripe ordering. The straight FS segments are also prone to produce (and originate from) nesting instabilities; that is, a large portion of the FS is connected in $\vec{k}$ space to another by the reciprocal lattice vector $\vec{q}$. Figure 4A shows the schematic plot explaining such an effect, where the main band hybridizes with an Umklapp band shifted by the amount $\vec{G}$ in $\vec{k}$-space. This hybridization opens a gap $2\Delta$ which lowers the electronic energy of the system. The heavy nesting we have observed means that large portions of the FS will be gapped, gaining more energy and making the distortion more likely. The fact that the entire FS can be affected by these vectors (Fig 4C) means that there will be no ungapped portions remaining, which is consistent with the insulating behavior above $T_c$.

The $\vec{q}$ vector that most naturally connects the straight FS segments is ($\pm$ 0.3,0) or (0,$\pm$0.3) in units of 2$\pi$/a (black lines in figure 4C). A charge density wave corresponding to this wave-vector would have a modulation period of 1/0.3 or 3.3 lattice constants in real space. This incommensurate CDW is illustrated in figure 4B. Such a distortion has in fact been recently observed in diffuse X-ray and neutron scattering measurements as a weak superlattice reflection with momentum transfer $\vec{q}_{orbital} = (0.3, 0, 1) = (\frac{0.6\pi}{a}, 0, \frac{2\pi}{c})$ *(3,4,5)*. These reflections are broad in $\vec{q}$-space, implying that they only exist with short range order, with a coherence length determined to be roughly 20 Å in the plane. This short range nature is expected, because the real-space lattice constant of the modulation (3.3a) is incommensurate with the lattice. An analysis of the intensities of 109 of these superlattice peaks has been carried out to determine the atomic displacements associated with this CDW*(5)*. What was found was a slowly modulated (period 3.3a) cooperative Jahn-Teller style distortion of the $MnO_6$ octahedra, with the strength of the distortion presumably matching the amplitude of the charge modulation (figure3B). Thus it is natural to believe that the ordering observed in these scattering measurements has as its origin the FS nesting uncovered here.

The coupling of the nesting-induced CDW with the Jahn-Teller distortion helps explain more of the details of the ARPES spectra. For example, the pseudogap depletes spectral weight from a large energy range (~ 1eV) around $E_F$, which is a very large energy scale for a CDW style gap alone. Further, the gap edges are experimentally observed to be "soft" or gently sloping, as opposed to the sharp edges expected from a simple CDW gap. The cooperation of the CDW with the Jahn-Teller distortion explains these unusual phenomena, as the typical energy scale of the Jahn-Teller distortion is near 1 eV. In addition, the X-ray results indicate that the magnitude of the Jahn-Teller distortion varies in space, meaning that the resulting gap will have various energy scales and so the gap edge will be soft. This also implies that the carrier hopping probability $t_{ij}$ will vary from one site to another. Recent theoretical calculations suggest that if such a variation in $t_{ij}$ exists, then nanoscale phase separation and percolative conduction are expected (39).

The ARPES data in Fig 4B, as well as those from optical experiments show that the pseudogap begins to fill in as the temperature is lowered through $T_c$. This is consistent with the weakening of the CDW ordering in the metallic phase. Although the scattering experiments cannot observe them in the ferromagnetic phase, a pseudogap is still observed to a lesser degree and the FS segments are still very straight. We take this as an indication of nanoscale fluctuating remnants of the ordering below $T_c$: They fluctuate violently in both space and time so that the scattering experiments are not able to observe them, but they nonetheless still very much affect the electronic structure.

An additional type of weak superlattice reflection is observed in the paramagnetic state in the X-ray scattering experiments on these crystals *(3,40)*. This is the so-called CE ordering and has a wave-vector $\vec{q}_{CE} = (\frac{1}{4}, \frac{1}{4}, 0) = (\frac{\pi}{2a}, \frac{\pi}{2a}, 0)$. This is the combination of antiferromagnetism and charge/orbital orderings and is a real space ordering similar to (but different than) a checker board. The CE vectors are drawn in figure 4C as the red arrows, and they are found to be only slightly shorter than the ideal nesting vector. This proximity to the nesting instability means that it can still partly gap the near-Fermi states, although presumably with less effectiveness than the (0.3,0,1) vector. This is consistent with the neutron scattering experiments, which found the CE superlattice peaks to fluctuate more dynamically than the (0.3,0,1) peaks *(5)*. However, at very high temperature where double exchange and the FS topology are less meaningful, the real-space CE ordering should stabilize as compared to the $\vec{k}$ space driven (0.3,0,1) ordering.

The results reported here have analogs in other important systems. For example, in high temperature superconducting cuprates $(La,Nd)_{2-x}Sr_xCuO_4$, recent ARPES experiments on the commensurately doped x=1/8 compound showed straight FS segments that are very reminiscent of what we have observed here*(41)*. Such straight Fermi segments have been interpreted as a type of nanoscale phase separation: 1D charge stripes in this system. At this doping level, the stripes are believed to be static and subsequently destroy the superconductivity. A more interesting topic of current debate is whether these 1D stripes exist dynamically at other doping levels, coexisting with or even possibly causing the superconductivity*(42)*. This dynamic phase separation is perhaps more similar to what we have observed in the layered manganites. Because short range fluctuating superlattice peaks are observed in the perovskite manganites as well *(43)*, these general properties will probably extend to those systems as well.

**CONCLUSION**

We have provided the first high energy/momentum resolution ARPES data on the layered CMR material $La_{1.2}Sr_{1.8}Mn_2O_7$, uncovering the FS as well as the key transport parameters $\lambda$ (mean free path), $m^*$ (effective mass), $v_F$ (Fermi velocity), $\tau$ (mean free time between scattering) and *n* (in-plane carrier density). With these we can calculate the in-plane conductivity which turns out to be too high unless we introduce the pseudogap effect at $E_F$ to explain either the low temperature poor metallic behavior or high temperature insulating behavior of these compounds. Evidence suggests that the pseudogap originates from a short range charge/orbital density wave enhanced by FS nesting. The CDW cooperates with the Jahn-Teller effect and competes with the itinerancy energy of double-exchange. This leads to nanoscale phase separation of the fluctuating charge ordered and metallic regimes – a key component to the CMR in these compounds *(44)*.

**REFERENCES AND NOTES**

1 J. Orenstein, A.J. Millis, Science **288**, 468 (2000)

2 C.D. Ling, *et al.*, Phys. Rev. B **62**, 15096 (2000)

3 L. Vasiliu-Doloc *et al.*, Phys. Rev. Lett **83**, 4393 (1999)

4 M. Kubota et al, J. Phys. Soc. Jpn, **69**, 1986 (2000)

5 B Campbell et al, in preparation.

6 Y. Moritomo, A. Asamitsu, H. Kuwahara, Y. Tokura, Nature **380**, 141 (1996)

7 C. Zener, Phys. Rev. **82**, 403 (1951)

8 P.-G. de Gennes, Phys. Rev. **118**, 141 (1960)

9 P. W. Anderson, H. Hasegawa, Phys. Rev. **100**, 675 (1955)

10 A.J. Millis, P.B. Littlewood, B.I. Shraiman, Phys. Rev. Lett. **74**, 5144 (1995)

11 Y. Tokura, *Colossal Magnetoresistive Oxides*, Advances in Condensed Matter Science (Gordon and Breach Publishers, 2000)

12 Z.X. Shen, D.S. Dessau, Phys. Rep. **253**, 1 (1995)

13 The ARPES experiments were performed at undulator Beamline 10.0.1 at the ALS using a Scienta SES 200 energy analyzer. All data shown here were taken with 50eV photons at 20K in a vacuum better than $4*10^{-11}$ torr. The angle mode of the analyzer was used, allowing us to simultaneously collect ~38 individual spectra along a ~12$^O$ angular slice in $\theta$. The energy resolution of the analyzer was better than 10 meV, measured from the 10 to 90% width of the Au reference taken at 10K and the angular resolution was $\pm 0.16^O$ along the angular slice and $\pm 0.25^O$ in the perpendicular direction $\phi$, giving a momentum resolution $\left(\Delta k_x, \Delta k_y\right) = \left(0.02\frac{\pi}{a}, 0.033\frac{\pi}{a}\right)$. To grid the Brillouin zone, the samples were rotated in the $\phi$ direction in 1$^O$ steps. The analyzer was left fixed with the central axis making an 83$^O$ angle relative to the photon beam. In this configuration, the 12$^O$ slices were parallel to the incident photon polarization direction which was mostly out of plane.

14 The EDCs presented here were normalized with emission intensity in a binding energy (0.1eV, 0.25eV) window. Such high order harmonic emission is isotropic and is proportional to photon flux.

15 N. Hamada, unpublished data.

16 D.S. Dessau *et al.*, Phys. Rev. Lett. **81**, 192 (1998)

17 S. Massidda et al., Physica C **152**, 251 (1988)

18 Y. -D. Chuang, *et al.*, preprint available at http://xxx.lanl.gov/abs/cond-mat/01021386

19 D.L. Feng *et al.* preprint available at http://xxx.lanl.gov/abs/cond-mat/0102385

20 T. Saitoh, *et al.*, Phys. Rev. B **62**, 1039 (2000)

21 The FS centered around (0,0) will of course have very different parameters, but these will not be very relevant for the in-plane conductivity.

22 Qing An Li, K. E. Gray, and J. F. Mitchell, Phys. Rev. B **63**, 24417 (2001)

23 A. Abrikosov, Phys. Rev. B **61**, 7770 (2000).

24 R. Joynt, Science **284**, 777 (1999).

25 D. S. Dessau, T. Saitoh, Science **287**, 767a (2000).

26 K Schulte *et al.,* PRB. 63, 165429 (2001)

27 T. Ishikawa, T. Kimura, T. Katsufuji, Y. Tokura, Phys. Rev. B **57**, R8079 (1998)

28 T. Ishikawa, K. Tobe, T. Kimura, T. Katsufuji, Y. Tokura, Phys. Rev. B **62**, 12354 (2000)

29 A.J. Millis, B.I. Shraiman, R. Mueller, Phys. Rev. Lett **77**, 175 (1996)

30 A.J. Millis, R. Mueller, B.I. Shraiman, Phys. Rev. B **54**, 5405 (1996)

31 D.J. Singh, W.E. Pickett, Phys. Rev. B **57**, 88 (1998).

32 H. Roder, J. Zang, A.I. Bishop, Phys. Rev. Lett. **76**, 1356 (1996).

33 A.S. Alexandrov and A.M. Bratkovsky, Phys. Rev. Lett **82**, 141 (1999).

34 V. Ferrari, M.J. Rozenberg and R. Weht, preprint available at http://xxx.lanl.gov/abs/cond-mat/9906131

35 A. Moreo, S. Yunoki, E. Dagotto, Science **283**, 2034 (1999),

36 A. Moreo, M. Mayr, A. Feiguin, S. Yunoki, E. Dagotto, Phys. Rev. Lett. **84**, 5568 (2000)

37 S. Mori et al., Nature (London) **392**, 473 (1998).

38 T. Hotta, A. Feiguin, E. Dagotto, Phys. Rev. Lett. **84**, 2477 (2000)

39 M. Mayr, *et al. ,* Phys. Rev. Lett. **86**, 135 (2001)

40 D. Argyriou et al, Phys. Rev B **61**, 15269 (2000)

41 X.-J. Zhou *et al.*, Science **286**, 268 (1999)

42 V.J. Emery, S.A. Kivelson, J.M. Tranquada, preprint available at http://xxx.lanl.gov/abs/cond-mat/9907228

43 S. Shimomura, N. Wakabayashi, H. Kuwahara and Y. Tokura,, , Phys. Rev. Lett. **83**, 4389 (1999)

44 We thank D. Argyriou, B. Campbell, R. Osborn, E. Dagotto, J. Zhang, A. Fedorov, and N. Hamada for helpful discussion; S. Kellar, X.J. Zhou, P. Bogdanov and Z. Hussain, for instrumentation help; and J. Denlinger for analysis software. Supported by NSF grant CAREER-DMR-9985492 and U.S. Department of Energy (DOE) grant DE-FG03-00ER45809. The ALS is supported by DOE.

**Fig 1.** (A) Doping phase diagram for $La_{2-2x}Sr_{1+2x}Mn_2O_7$, extracted from Ling et al. *(2)*. (B) Resistivity versus temperature curves for $La_{1.2}Sr_{1.8}Mn_2O_7$ at various magnetic fields, after Moritomo *et al.(6)*.

**Fig 2.** Low temperature ($T = 20K$) ARPES data from $La_{1.2}Sr_{1.8}Mn_2O_7$: (A) The integrated spectral weight near $E_F$ over much of the first Brillouin zone, and the LSDA FS (black lines). The data have been reflected about the $(0,0)-(\pi,0)$ line. (B) Binding energy versus emission angle ($\theta$) image plot from an angular slice centered at $(0, 0.7\pi)$ [see the horizontal white line in (A)]. (C) Near-$E_F$ blow up of data from (B)[blue box], with the color rescaled. (D,E) Approximately one-third of the EDCs from (B) and (C). Corresponding angles are listed in (E). (F) Overlay of the EDC at $4.67^o$ with a Au reference spectrum. (G) MDCs from (C) for every 10 meV from binding energy -0.2 eV to 0.02eV.

**Fig 3**. Fitted results from MDCs: (A) Spectral weight (red curve) under red branch in (B) versus binding energy. The black dotted line indicates the expected weight behavior for a non-interacting theory. (B) $\vec{k}$-centroids (red and blue) and $\vec{k}$-half width (green) versus binding energy. (C) Temperature dependence of the weight remaining at $E_F$, after Saitoh et al (20).

**Fig 4** (A) Schematic plot for density wave formation – k-space. (B) Density wave formation in real space with a lattice constant 3.3a. (C) Overlay of experimental FS topology with CE type ordering vectors (red arrows), and orbital stripe ordering vectors (black arrows)

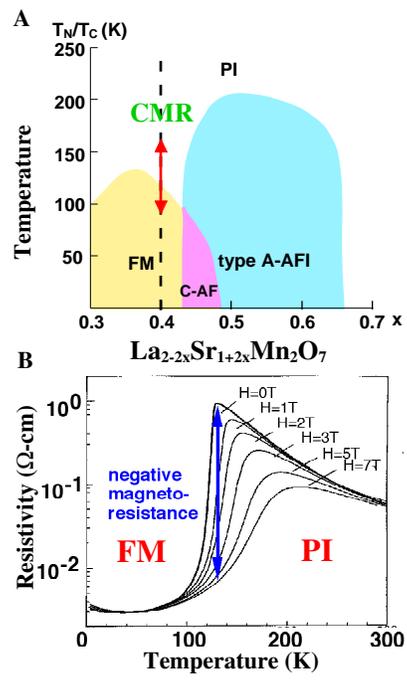

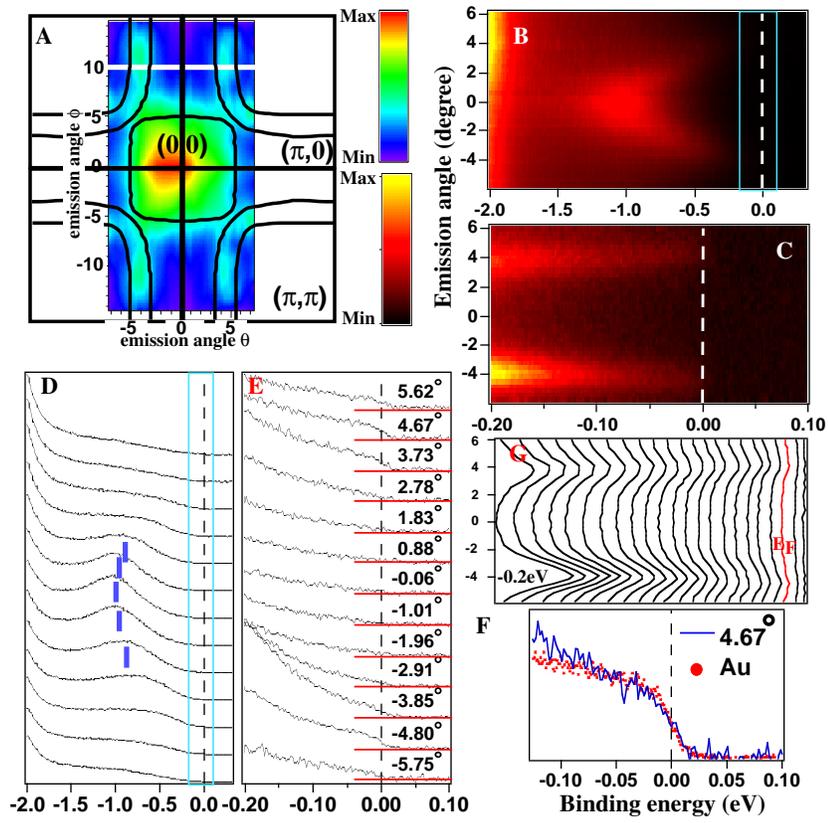

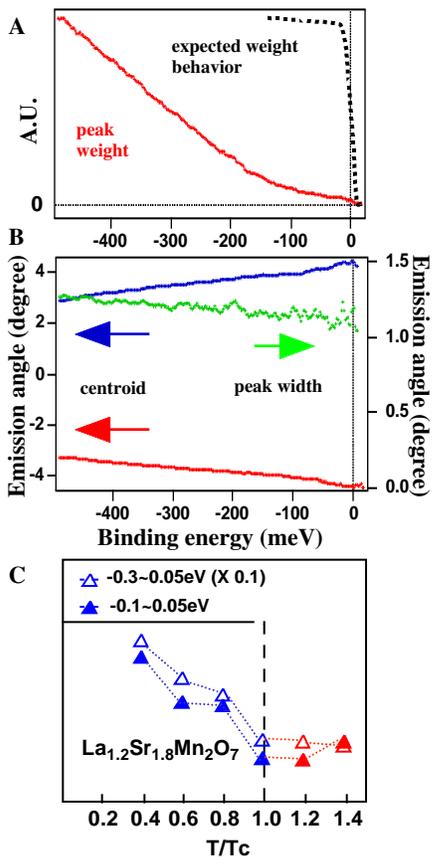

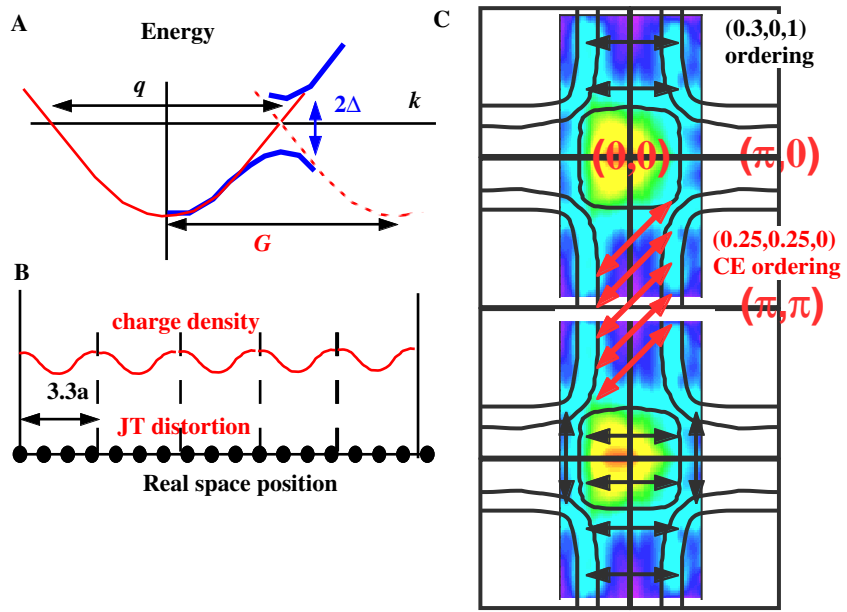